\documentclass[twocolumn,showpacs,superscriptaddress,secnumarabic,amsmath,nobibnotes,bm,pre,prb]{revtex4-2}
\usepackage{dcolumn}
\usepackage{bm}

\usepackage{verbatim}

\usepackage{color}
\usepackage[hypertexnames=false]{hyperref}
\usepackage{tabularx}
\usepackage{amsmath}
\usepackage{siunitx}
\usepackage{comment}
\usepackage{xcolor}
\usepackage{graphicx}
\usepackage{siunitx}
\usepackage{adjustbox}

\usepackage{multirow}
\usepackage{tcolorbox} 
\usepackage{booktabs}  
\usepackage{array}   
\usepackage{colortbl} 
\usepackage{xfrac}
\usepackage{booktabs}

\definecolor{customgreen}{HTML}{c7efcf}
\definecolor{customblue}{HTML}{9cafb7}

\newcommand{\be}{\begin{equation}}
\newcommand{\ee}{\end{equation}}
\newcommand{\bea}{\begin{eqnarray}}
\newcommand{\eea}{\end{eqnarray}}
\newcommand{\bei}{\begin{itemize}}
\newcommand{\eei}{\end{itemize}}

\begin{document}
\author{Mohammad Aghaie}
\affiliation{Dipartimento di Fisica ``E. Fermi", Università di Pisa, Largo Pontecorvo 3, 56127 Pisa, Italy}
\affiliation{INFN Sezione di Pisa, Polo Fibonacci, Largo B. Pontecorvo 3, 56127 Pisa, Italy}

\author{Pedro De la Torre Luque}
\affiliation{Departamento de Física Teórica, M-15, Universidad Autónoma de Madrid, E-28049 Madrid, Spain}
\affiliation{Instituto de Física Teórica UAM-CSIC, Universidad Autónoma de Madrid, C/ Nicolás Cabrera, 13-15, 28049 Madrid, Spain}

\author{Alessandro Dondarini}
\affiliation{Dipartimento di Fisica ``E. Fermi", Università di Pisa, Largo Pontecorvo 3, 56127 Pisa, Italy}
\affiliation{INFN Sezione di Pisa, Polo Fibonacci, Largo B. Pontecorvo 3, 56127 Pisa, Italy}
\affiliation{Galileo Galilei Institute for Theoretical Physics, Largo Enrico Fermi 2, I-50125 Firenze,
Italy}

\author{Daniele Gaggero}
\affiliation{INFN Sezione di Pisa, Polo Fibonacci, Largo B. Pontecorvo 3, 56127 Pisa, Italy}

\author{Giulio Marino}
\affiliation{Dipartimento di Fisica ``E. Fermi", Università di Pisa, Largo Pontecorvo 3, 56127 Pisa, Italy}
\affiliation{INFN Sezione di Pisa, Polo Fibonacci, Largo B. Pontecorvo 3, 56127 Pisa, Italy}

\author{Paolo Panci}
\affiliation{Dipartimento di Fisica ``E. Fermi", Università di Pisa, Largo Pontecorvo 3, 56127 Pisa, Italy}		
\affiliation{INFN Sezione di Pisa, Polo Fibonacci, Largo B. Pontecorvo 3, 56127 Pisa, Italy}

\bibliographystyle{apsrev4-1}
	
  \title{(H)ALPing the 511 keV line:\\ A thermal DM interpretation of the 511 keV emission}	
  \begin{abstract}
We propose a novel framework where MeV-scale Dirac Dark Matter annihilates into axion-like particles, providing a natural explanation for the 511 keV gamma-ray line observed in the Galactic Center. The relic abundance is determined by $p$-wave annihilation into two axion-like particles, while $s$-wave annihilation into three axion-like particles, decaying into $e^+e^-$ pairs, accounts for the line intensity. Remarkably, this model reproduces the observed emission morphology, satisfies in-flight annihilation and cosmological bounds, and achieves the correct relic density, assuming a contracted Navarro-Frenk-White profile similar to the one that reproduces the Fermi-LAT Galactic Center Excess, offering a compelling resolution to this longstanding anomaly.
  \end{abstract}
\maketitle
 
A prominent spectral line at 511 keV has been identified in observations of diffuse $\gamma$-ray emission at the MeV energy scale~\cite{1972ApJ...172L...1J, 1973ApJ...184..103J, 1986ApJ...302..459L}. This emission is attributed to the annihilation of electrons and positrons forming a para-positronium (p-ps) bound state. However, the observed emission cannot be fully explained by positrons produced through cosmic-ray (CR) interactions with the interstellar medium (ISM). This discrepancy suggests evidence of an additional steady positron injection from sources located within in the inner Galaxy, with a high statistical significance~\cite{Siegert:2015knp}. 
In addition to the 511 keV line, a continuum emission is present in the photon flux, primarily due to: \(i)\) the decay of the triplet ortho-positronium (o-ps) configuration, \(ii)\) in-flight annihilation (IfA) from the direct annihilation of positrons with ambient electrons, and \(iii)\) the Inverse Compton (IC) scattering of CR electrons off Galactic radiation fields. These observations indicate that positron sources significantly contribute to the diffuse $\gamma$-ray flux around $\sim\,\si{MeV}$ energies~\cite{Bandyopadhyay2009, Stecker}. 
Morphological studies of this signal outlined a concentrated {\it bulge} emission and a weaker, more extended {\it disk} emission.
Various sources have been proposed in the literature, among which $\beta^+$-emission from radionuclides produced in massive stars appears to account for the most of the latter ~\cite{RevModPhys.83.1001, Kn_dlseder_2005, Weidenspointner:2006nua}. 
However, the spatial morphology and injection rate of positrons in the bulge are inconsistent with known candidates, unless a unique source population with a highly concentrated central distribution is considered~\cite{Bartels:2018eyb}. 
 \begin{figure}[t!]
    \centering
    \includegraphics[width=0.95\linewidth]{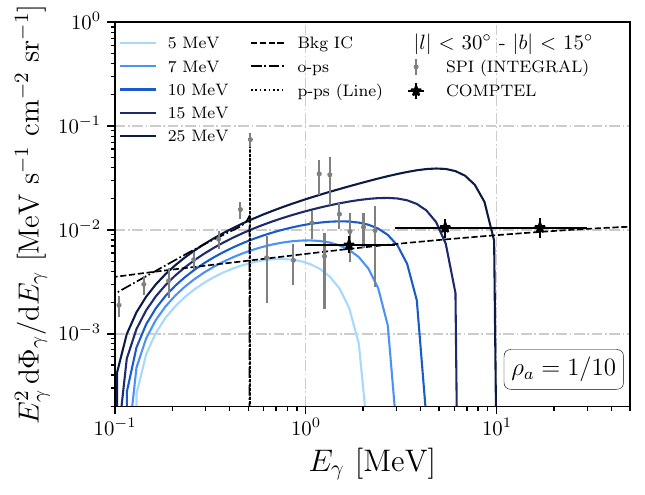}
    \caption{The IfA emission spectrum associated with the $e^+$ injection from DM annihilation in our model, is presented. We consider different DM masses $m_\chi$ above $10\,\mathrm{MeV}$ and compare the resulting IfA signals with observations of the MeV $\gamma$-ray diffuse Galactic emission. The analysis focuses on a specific region of interest, defined as $|l|<30^\circ$ and $|b|<15^\circ$, where observational data from SPI (aboard INTEGRAL)~\cite{Bouchet:2011fn} and COMPTEL (aboard CGRO)~\cite{COMPTEL1994} are available. More details in the text.
    } 
    \label{fig:IfA_MassScan}
\end{figure}

\smallskip
An appealing solution was suggested by Ref.~\cite{Boehm:2003bt}: MeV dark matter (DM) directly annihilating into $e^+e^-$ pairs. 
One of the main motivation is that the morphology of the 511 keV emission fit quite well with the square of a Navarro-Frenk-White (NFW) DM density profile~\cite{Navarro_1996}. This means that, while decaying DM fails to reproduce the correct spatial morphology of the 511 keV line~\cite{Boehm_2004}, velocity-independent DM annihilation is consistent with the observed bulge emission. 

On a more specific level, to explain the Galactic Center signal, a model must satisfy three key requirements: 
\textit{i)} accurately predict the total flux, 
\textit{ii)} reproduce the line spectrum, and 
\textit{iii)} match the observed spatial morphology of the emission. 
For annihilating DM, with mass $m_{\chi}$, directly into $e^{\pm}$ pairs, achieving the required total flux demands $\langle\sigma v\rangle/m_\chi^2 \sim 10^{-31}\,\si{cm^3\,s^{-1}\,MeV^{-2}}$
(see Refs.~\cite{Wilkinson_2016, Boehm:2003bt}). Matching the line spectrum requires a DM mass below $10\,\si{MeV}$ to avoid constraints from IC scattering and IfA. In particular, Ref.~\cite{Beacom_2006} demonstrated that the continuum emission from positrons capable of explaining the 511 keV signal would exceed the diffuse Galactic photon flux in the MeV range if these positrons were injected with energies above a few MeV ($\sim 3 \, \si{MeV}$). Thus, in the simplest scenario where DM annihilates directly into $e^\pm$ pairs, this translates into an upper limit on the DM mass. Ref.~\cite{sizun2007constraintsinjectionenergypositrons} revisited this analysis, softening the upper bound to approximately $7\,\si{MeV}$, but certainly below $10\,\si{MeV}$. These combined constraints on $\langle\sigma v\rangle$ and $m_\chi$ significantly restrict the parameter space, often leading to $\langle\sigma v\rangle \approx 10^{-31}\,\si{cm^3\,s^{-1}}$.
Such a small cross-section results in either a severe overproduction of DM in the early universe, which may potentially be avoided in models with leading $p$-wave annihilations, or necessitates additional annihilation channels into light products. These additional channels, however, risk conflicting with recombination and Big Bang Nucleosynthesis (BBN) constraints (see Ref.~\cite{Sabti_2020}). 
On top of this, as noted in~\cite{DelaTorreLuque:2023cef}, a NFW profile alone is insufficient to reproduce the observed 511 keV longitude profile when including the propagation of positrons injected with energies above $\sim10$~MeV.

As a result, the straightforward explanation of thermal DM directly annihilating into $e^\pm$ pairs has lost hype and is largely disregarded as a viable source of the excess. Additionally, more exotic DM models have been proposed, such as excited DM~\cite{Pospelov, Cappiello_2023}, but none have yet provided a solid and natural explanation for the observed excess.

\medskip
In this work, we present a novel theoretical framework in which a thermally produced light  DM candidate annihilates into axion-like particles (ALPs). The subsequent decay of ALPs produces multiple $e^\pm$ pairs, significantly reducing the kinetic energy of the injected positrons. This distinctive kinematic behavior permits DM masses exceeding 10 MeV while avoiding violations of constraints from the Galactic diffuse MeV flux. Furthermore, the model naturally accounts for both the normalization and morphology of the observed 511 keV line, remaining fully consistent with all cosmological and astronomical constraints. This provides a robust and compelling solution to a long-standing astrophysical problem.

\medskip
\emph{Theoretical Framework ---} 
 The model under consideration extends the Standard Model (SM) by introducing an ALP \(a\), which couples to all three SM leptons \(\ell = e, \mu, \tau\) and to a Dirac DM particle \(\chi\). Consequently, the relevant interaction Lagrangian is given by 

\begin{equation}  
   \mathcal{L}_{\rm int} \supset- i a \left( g_{\chi} \bar{\chi} \gamma_5 \chi + g_{\ell} \bar{\ell} \gamma_5 \ell \right) \ , 
   \label{eq:lagrangian}
\end{equation}  
assuming that the couplings \(g_{\chi}\) and \(g_{\ell}\) are real. 

\smallskip
We consider the mass hierarchy $2m_\chi>3m_a>6m_e$ where $m_\chi<100$\,MeV, which ensures that DM can produce ALPs that subsequently decay only into $e^\pm$ at tree level and photons at loop level. As a result, the coupling to electrons $g_e$ is the primary focus. In this mass hierarchy, the relevant annihilation processes proceed via the following t-channel mechanisms:
\begin{itemize}
    \item[$i)$] the \(p\)-wave: \(\chi \bar{\chi} \rightarrow aa\),
    \item[$ii)$] the \(s\)-wave: \(\chi \bar{\chi} \rightarrow aaa\).
\end{itemize}

These two processes are entirely governed by the coupling $g_\chi$, while the coupling $g_e$ only affects the ALP mean-free path. If the mean-free path is smaller than the parsec scale, the two t-channel processes result in an effectively instantaneous injection of $e^\pm$ from the perspective of the 511 keV line, with the cross section entirely determined by $g_\chi$. Conversely, direct annihilation into $e^\pm$ via off-shell ALP mediation does depend on the electron coupling $g_e$. However, since $g_e$ is strongly constrained, as will be discussed later, the annihilation cross section into $e^\pm$ pair via off-shell ALP mediation contributes negligibly.

\smallskip
In the non-relativistic limit, $s\simeq m_\chi^2(4 +\bar{v}^2)$, where $\bar{v}$ is the DM relative velocity, the first process provides a cross section that takes the form~\cite{Arina:2014yna, Armando_2024}:

\begin{equation}
   \sigma v_{\chi \chi\to aa} =  \frac{g_\chi^4 \bar{v}^2}{24\pi} 
  \frac{m_\chi^2(m_\chi^2 - m_a^2)^2}{(2m_\chi^2 - m_a^2)^4} 
 \sqrt{1 - \frac{m_a^2}{m_\chi^2}} \ .
\label{eq:xsecaa}
\end{equation}

Regarding the second process, the cross section simplifies in the massless limit, $m_a \to 0$, reducing to~\cite{Kahlhoefer_2017}:
\begin{equation}
    \sigma v_{\chi\chi\to  aaa} \simeq \frac{(7\pi^2 - 60) g_\chi^6}{1536 \pi^3 m_\chi^2}\ .
\label{eq:xsecaaa}
\end{equation}
The more general expression with massive ALPs is given in Appendix~\ref{sec:ann}.

\smallskip
The two processes above are relevant at different stages of cosmological evolution. The $p$-wave channel is crucial for determining the relic density as during freeze-out the relative velocity is sufficiently high. On the other hand, the $s$-wave process is relevant at late time. In the massless limit, the ratio of the cross sections takes the form:
\begin{equation}
   \eta\equiv \frac{ \sigma v_{\chi\chi\to  aaa}}{ \sigma v_{\chi\chi\to  aa}} \approx 0.23 \, \frac{g_\chi^2}{\bar{v}^2} \ .
  \label{eq:xsecsratio}
\end{equation} 
In low-velocity environments, such as the Galactic bulge, both the phase-space and coupling suppressions in Eq.~\eqref{eq:xsecaaa} are alleviated. Specifically, considering the uncertainty in the dispersion velocity of DM particles in the Galactic bulge, ranging from  $ 50\,\si{km\,s^{-1}}$ to $ 140\,\si{km\,s^{-1}}$~\cite{DiskVel1, DiskVel2, DiskVel3}, and $g_\chi$ values between $10^{-3}$ and $10^{-2}$ (consistent with thermal production, as discussed later) there exists a reasonable velocity choice such that the ratio \(\eta \gg 1\) today. Consequently, the 511 keV line emission is predominantly driven by the  \(s\)-wave annihilation process into three ALPs.

\begin{figure*}[t!]
    \centering
    \includegraphics[width=0.47\linewidth]{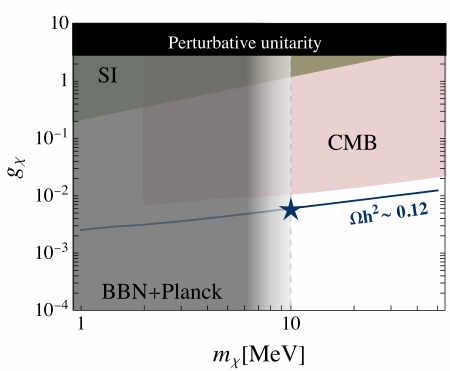} \
    \includegraphics[width=0.455\linewidth]{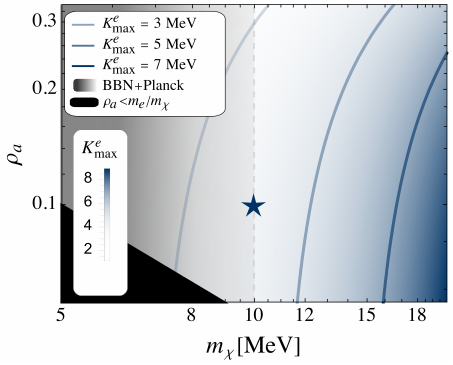}
    \caption{
    \textbf{Left Panel:} 
   Parameter space in the $(m_\chi, g_\chi)$ plane, with the red, green, and black shaded regions representing the CMB, SI, and perturbative unitarity constraints, respectively. The solid blue line shows the thermal relic, while the blue star marks the thermal benchmark point with $m_\chi = 10\,\text{MeV}$ and $\rho_a = 1/10$. The vertical dashed gray line indicates the $2\sigma$ constraint from BBN+Planck data~\cite{Sabti_2020}, and the gray gradient highlights regions in tension with cosmological constraints at the $3\sigma$ level (low masses).
    \textbf{Right panel:}
   Parameter space in the $(m_\chi, \rho_a)$ plane, with the maximum $e^+$ kinetic energy shown by an increasing blue gradient: darker blues correspond to higher values of $K_{\rm max}^{e}$. The three curves represent fixed values of $K_{\rm max}^{e} = 3, 5, 7\,\text{MeV}$. The vertical dashed gray line, gray gradient, and blue star are the same as in the left panel. The black shaded area represents the forbidden region where $m_a < 2m_e$.
    }
    \label{fig:constraints}
\end{figure*}

\medskip
For this model we have taken into account the following constraints on both $g_e$ and $g_\chi$:
\begin{itemize}
\item[$\diamond$]\textbf{Thermal freeze out:} 
We consider the scenario in which the dark sector decouples from the SM bath at a temperature $T_{a\rm SM}$ above the electroweak scale and therefore well before freeze-out. As a consequence, the dark sector temperature $T'$ becomes lower than the SM plasma temperature $T$, according to the ratio of the effective degrees of freedom: $T' = (g_\star(T)/g_\star(T_{a\rm SM}))^{1/3} T$.
In this situation, since dark-sector particles remain in thermal equilibrium (for sufficiently large $g_\chi$), freeze-out proceeds within the dark sector at temperature $T'$. We solve the Boltzmann equation for thermal freeze-out considering only the dominant $p$-wave annihilation channel $\sigma v_{\chi\chi\to aa}$. The thermally averaged annihilation cross section in the dark sector is obtained by replacing $\bar{v}^2$ with $6/x'$, where $x' = m_\chi / T'$. By requiring the relic abundance $\Omega h^2 \simeq 0.12$~\cite{Planck2018}, we determine the coupling $g_\chi$ as a function of the DM mass; the result is shown as the solid blue line in the left panel of Fig.~\ref{fig:constraints}. From this analysis, we find $g_\chi \simeq 2.5 \times 10^{-3} (m_\chi / \mathrm{MeV})^{1/2}$ with only a mild dependence on $m_a$ within the parameter space explored in this work; see Appendix~\ref{sec:relic density} for details.

\item[$\diamond$] \textbf{ALP-DM coupling}: Cosmological bounds constrain the ALP-DM coupling $g_\chi$ and we show the results in the left panel of Fig.~\ref{fig:constraints}. In particular data from BBN and Planck on $\Delta N_{\text{eff}}$ impose strong limits on $m_\chi$, with~\cite{Sabti_2020} finding $m_\chi \lesssim 10$\,MeV at $2\sigma$ confidence level (CL) for Dirac DM particle. The excluded region is represented by the gray shaded region. The CMB provides additional constraints, as DM annihilations inject energy into the plasma~\cite{Peebles:2000pn, Chen:2003gz, PhysRevD.72.023508}.
For the positron annihilation channel, we adopt the stringent bounds from~\cite{PhysRevD.93.023527} and the exclusion is shown as a red shaded region. Other constraints arise from the perturbative unitarity limit (black shaded region), $g_\chi < \sqrt{8\pi/3}$~\cite{Cornella_2020}, and DM self-interactions (SI) (green shaded region), which impose \(g_\chi \lesssim 0.21 \left( {m_\chi}/{\text{MeV}} \right)^{3/4}\)~\cite{PhysRevD.87.115007, Spergel_2000}.

\item[$\diamond$] \textbf{ALP-electron coupling}: ALPs that couple to electrons at tree-level face various constraints discussed extensively in the literature~\cite{ALPBoundGen1, ALPBoundGen2, ALPBoundGen5, SNBound, BBNALP1, BBNALP2, BeamDumpE137-1, BeamDumpE137-2}. In the parameter space of interest, the most stringent limits arise from supernova observations, including constraints from excessive cooling and the absence of gamma-ray bursts associated with ALP decay~\cite{SNBound}. Additional constraints are provided by beam dump experiments~\cite{BeamDumpE137-1, BeamDumpE137-2}. When these constraints are combined, for ALP masses in the MeV to tens of MeV range, the allowed window for the electron coupling is $10^{-11} \lesssim g_e \lesssim 10^{-9}$, implying that \(g_e \ll g_\chi\). A recent study on supernova production of ALPs~\cite{Fiorillo:2025sln} has updated these bounds, demonstrating that this window is now closed by constraints from energy deposition in supernovae. The analysis now restricts the electron coupling to $g_e \lesssim 5\times 10^{-13}$ for ALP masses in the MeV to tens of MeV range. Within this more constrained range of couplings, we find: \textit{i)} DM annihilation via off-shell ALP mediation is entirely negligible, and \textit{ii)} the ALP mean-free path is smaller than the parsec scale, ensuring an effectively instantaneous $e^\pm$ injection for the 511 keV line signal. and \textit{ii)} the ALP mean free path is smaller than the parsec scale, ensuring an effectively instantaneous $e^\pm$ injection for the 511~keV line signal. Indeed, in the relevant parameter space the average ALP mean free path is $\lambda_a = \beta_a \gamma_a \, \Gamma_a^{-1}$, with $\Gamma_a \propto g_e^2 m_a$. For $g_e \simeq 10^{-14}$ and $m_a \sim \mathcal{O}(\mathrm{MeV})$, this corresponds to $\lambda_a \lesssim 1~\mathrm{pc}$. However, we note that with such a small couplings, the ALP lifetime is long enough to have an impact in the early early Universe cosmology, in particular for BBN. While a thorough study of this issue is beyond the scope of the current letter, it represents an important topic for future work.
\end{itemize}

As can be seen, this model is well-motivated because it allows for a thermal candidate that is not yet excluded by cosmological constraints and, as we show, reproduces the morphology of the 511 keV line.

\medskip
\emph{Kinematics ---} 
The kinematics of this model, driven by the chain $\chi \bar\chi \to 3a \to 3(e^+e^-)$, represents a distinguishing feature compared to other DM models previously studied in the literature. The key-point of this scenario is that DM generates $e^\pm$ with energies substantially lower than its mass. 
By defining \(\rho_a = m_a / (2m_\chi)\), the mass hierarchy of interest corresponds to \(\rho_a \in \left(m_e / m_\chi, 1/3\right)\).
The differential cross-section of the injected positrons can be determined by convolving the differential cross-section for the process $\chi\bar \chi \to 3a$ with the energy spectrum of $e^\pm$ resulting from the two-body decay of each ALP. We approximate the $e^\pm$ injection spectrum using a Dirac delta function centered at half of the ALP energy. 
Defining $x_{1,2} = E_{1,2}/2m_\chi$, where $E_{1,2}$ are the energies of two of the three pseudoscalars in the center-of-mass frame of the DM pair, the $e^\pm$ injection rate per unit volume and energy, resulting from the DM annihilation process, can be expressed by the source term: 
\begin{equation}
    Q_e(\vec{x},E_e)=\frac{1}{4}\bigg(\frac{\rho_\chi(\vec{x})}{m_\chi}\bigg)^2 \left. \frac{1}{m_\chi} \frac{d \sigma v_{\chi \chi\to aaa}}{d x_2} \right|_{x_2 = \frac{E_e}{m_\chi}} \ ,
\end{equation}
where further details over $d  \sigma v_{\chi \chi\to aaa}/d x_2$ can be found in Appendix~\ref{sec:ann}. Here, we adopt a generalized NFW profile~\cite{gNFW} for the DM density, $\rho_\chi(\vec{x})$, fixing the inner and outer slopes to $\alpha = 1$ and $\beta = 3$, respectively, and allowing the contraction index $\gamma$ to vary. We take a scale radius $R_s = 20~\mathrm{kpc}$, with the normalization fixed by requiring $\rho_\chi(r_\odot)=0.4~\mathrm{GeV/cm^3}$.

\smallskip
The positron energy spectrum is entirely governed by the parameter $x_2$, which is kinematically constrained between $\rho_a$ and $(1-\rho_a)/2$. As a consequence, the kinetic energy range of the injected positrons is determined by  
\begin{equation}
K_{\rm min}^{e}= \rho_a m_\chi - m_e \ , \qquad K_{\rm max}^{e}= \frac{1-\rho_a}2 m_\chi - m_e \ .
\end{equation} 
From these equations, it follows that as $\rho_a \to 1/3$, the source term $Q_e$, as a function of the $e^+$ kinetic energy, becomes sharply peaked around $K_{e}=m_\chi/3-m_e$. Conversely, as $\rho_a \to m_e/m_\chi$, $Q_e$ broadens, with $K_{e}$ spanning from zero to $\left(m_\chi-3m_e\right)/2$. Furthermore, it is important to point out that the total cross section into three ALPs decreases as $\rho_a \to 1/3$, reaching its maximum value when $\rho_a \to m_e/m_\chi$.  

\smallskip
In the right panel of Fig.~\ref{fig:constraints} we explore the allowed parameter space in the $(m_\chi, \rho_a)$ plane, with the maximum $e^+$ kinetic energy shown by an increasing blue gradient. For reference, the three curves represent three fixed values of $K_{\rm max}^{e} = 3, 5, 7\,$MeV. As in the left panel, the gray shaded regions indicate the Planck+BBN constraint and the black shaded region is kinematically forbidden by ALP kinematics.

We adopt hereafter $\rho_a = 1/10$ and DM masses of $10\,\si{MeV}$ (blue stars in Fig.~\ref{fig:constraints}) and above.
This choice ensures compliance with IfA constraints while remaining consistent with the BBN and CMB constraints. A larger scan of the parameter space is provided in Appendices~\ref{sec:IfA_App} and~\ref{sec:Line_App}.

\medskip
Having at our disposal the source term, we can now compute the IfA emission, upgrading the procedure described in Ref.~\cite{Beacom_2006} for a continuum energy spectrum at injection. This estimation assumes that the IfA signal is generated from the positrons at their injection energy and its normalization only depends on the observation of the $511$~keV line by SPI, making it independent to any other astrophysical assumptions.
We note that the evaluation of the IfA emission follows the general form discussed in Ref.~\cite{Beacom_2006}, instead of that implemented in Ref.~\cite{DelaTorreLuque:2024wfz}. The reason is that the procedure from Ref.~\cite{Beacom_2006} is more conservative and totally model independent, besides to the fact that it offers a very simple way to show that a DM model that reduces the energy of the injected positrons leads to a significantly lower IfA emission associated. Meanwhile, the procedure from Ref.~\cite{DelaTorreLuque:2024wfz} assumed that the IFA signal is generated by the steady-state diffuse positrons, whose energy distribution is different to the injected one because of energy losses, reacceleration and other processes related to the propagation of the positrons, and, hence, less general.

In Fig.~\ref{fig:IfA_MassScan}, we show, for different DM masses, the expected IfA emission (solid lines) in a given region of interest (longitude $|l|<30^\circ$ and latitude$\,|b|<15^\circ$), where the SPI (gray points) and COMPTEL (black points) data are available. Here 
the overall normalization is adjusted to reproduce the intensity of the p-ps (dotted line) signal~\cite{Beacom_2006} and the associated o-ps continuum (dot-dashed line). We also show the IC predicted background emission from CR electrons (dashed line), derived from Refs.~\cite{delaTorreLuque:2022vhm, DelaTorreLuque:2023zyd}. 
For full details, we refer the reader to the discussion around Eq.~(2.13) of Ref.~\cite{DelaTorreLuque:2024wfz}.
We remark that low-energy data from SPI are affected by large systematic errors due to the subtraction of a relevant contribution from unresolved point sources, as discussed in Ref.~\cite{DelaTorreLuque:2024wfz}. Therefore, the mild overshooting of the low-energy continuum (which includes o-ps continuum, IC and IfA) does not affect our interpretation of the line signal itself. We remark that: \textit{i)} as we will show, this data-driven normalization is naturally obtained for thermal coupling and a contracted NFW DM profile similar to the one that reproduces the Fermi-LAT Galactic Center Excess~\cite{Ackermann_2017}; and \textit{ii)} the IfA spectrum for a DM particle of mass around 7-12 MeV exhibits a cutoff at around $E_{\gamma}=3$~MeV, which aligns well with the COMPTEL signal revealed in Ref.~\cite{Knodlseder_2025} that shows a strong correlation with the 511 keV excess. This DM scenario thus provides a compelling explanation for both excesses while remaining consistent with cosmological constraints.

Our results indicate that the expected IfA signals are compatible with data up to a mass of approximately $15$~MeV. 
Introducing an additional $30\%$ systematic uncertainty to the data, as performed in Ref.~\cite{Beacom_2006}, would significantly relax this upper limit. 
We also notice that the continuum $\gamma$-ray emission from the $e^{\pm}$ produced by DM—not only the IfA emission but also the bremsstrahlung emission—could contribute to explain the apparent excess in the diffuse MeV emission observed in the inner Galaxy, as highlighted in several studies~\cite{Karwin_2023, Orlando_2017, DelaTorreLuque:2024wfz, DelaTorreLuque:2024zsr}.

\medskip
In summary, the key feature of this thermal DM candidate is that positrons are injected with low energy, resulting in low IfA emission and a cutoff at a few MeV,. This allows for higher DM masses to remain compatible with the measurements of the Galactic $\gamma$-ray diffuse emission and the recent COMPTEL signal~\cite{Knodlseder_2025}.
 
\medskip
 \emph{Line Morphology---} As a final step, we demonstrate that this model can reproduce the latitude ($b$) and longitude ($l$) profiles of the signal, without requiring additional non-standard astrophysical assumptions. 

\smallskip
In order to obtain the line signal, we simulate the propagation of the low-energy leptons originating from DM annihilation by means of the {\tt DRAGON2} code \cite{Evoli:2016xgn}. This package is designed to compute all the relevant processes associated to the transport of charged particles in the Galaxy in a wide energy range, including diffusion, stochastic re-acceleration, advection, spallation and all the relevant energy losses. 
Our propagation setup adopts the transport parameters discussed in Ref.~\cite{DelaTorreLuque:2023olp}, which are derived from state-of-the-art analyses of secondary CRs in the Galaxy. More details on this computation are provided in Appendix~\ref{sec:Line_App}. We then follow the detailed prescription of Ref.~\cite{DelaTorreLuque:2023cef} (see also Ref.~\cite{DelaTorreLuque:2024wfz}) to compute the 511 keV emission from the steady-state distribution of positrons obtained with {\tt DRAGON2}. 

With this result at hand, we can directly compare with the SPI observations of the spatial profiles of the emission~\cite{Siegert:2015knp}. 
We note that these measurements are highly dependent on the templates used to derive the data, so that extra systematic uncertainties, not well estimated and not shown in the data, are also present. However, we do not aim to give the best-fit choice of DM parameters, but to show that this DM candidate provides a satisfactory agreement with the data.

\begin{figure}[t!]
    \centering
    \includegraphics[width=1\linewidth]{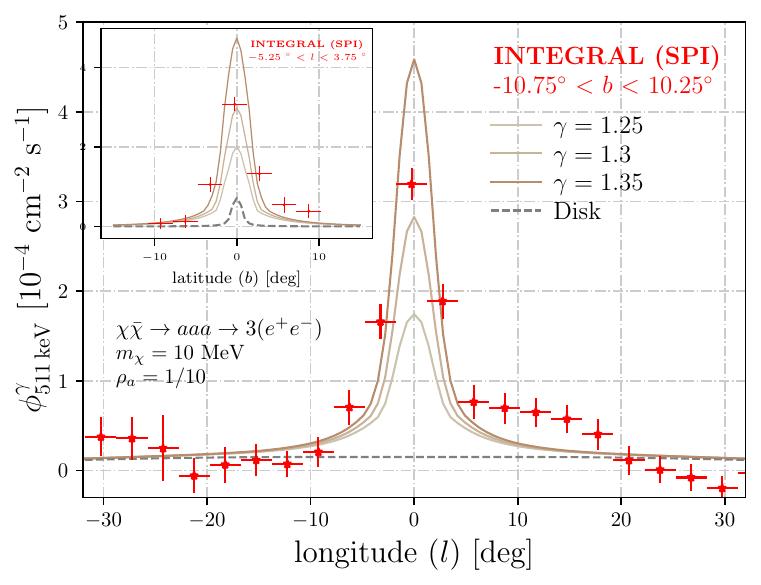} 
    \caption{Predicted longitude and latitude (inset panel) profiles for different values of the slope ($\gamma$) which represent the contraction of the DM distribution. These results correspond to the benchmark point of $m_\chi = 10$~MeV and $\rho_a = 1/10$, represented by the blue stars in Fig.~\ref{fig:constraints}. The disk model is taken from \cite{Robin:2003uus}.
    }
    \label{fig:511keV_profile}
\end{figure}

Fig.~\ref{fig:511keV_profile} shows the predicted longitude and latitude (inset panel) profiles of the 511 keV emission from annihilation of our DM candidate at a mass of $10$~MeV and $\rho_a = 1/10$, for a generalized NFW DM distribution with $\gamma$ ranging from 1.25 to 1.35, which is, interestingly, consistent with the Fermi-LAT Galactic Center Excess (indeed, the possible connection between both excesses have been proposed in the past~\cite{Bartels_2018, Silk_2018}). We note that reasonable variations in the density normalization and scale radius of the DM profile do not qualitatively affect this conclusion, as they primarily rescale the overall signal normalization and induce only mild changes to the morphology in the inner Galactic region. Moreover, the expected contribution from a young stellar disk, as mentioned in the Introduction, is also taken into account, following the parametrization of \cite{Robin:2003uus,Vincent:2012an}. The parameters of the disk model are chosen as in \cite{DelaTorreLuque:2024wfz}.
Remarkably, the predicted latitude and longitude profiles are able to reproduce the observed bulge-to-disk ratio for a DM distribution very close to a NFW, with no need of assuming any DM spike~\cite{DelaTorreLuque:2024wfz}. 
We show the same profiles for different masses and values of \(\rho_a\) in Fig.~\ref{fig:App_511keV_rhos} of Appendix~\ref{sec:Line_App}. As we see from these figures, different combinations of $\rho_a$ and the contraction index of the DM distribution allow to reproduce the profiles also for masses above $10$~MeV. Moreover, we notice that different combinations of $\rho_a$, DM mass and DM profile would require different levels of contribution to the 511 keV line from astrophysical sources (e.g. massive stars) to explain the disk emission. 

\emph{Conclusions ---}  In this letter we introduced a novel theoretical framework where a thermally produced DM candidate annihilates into ALPs. The subsequent decay of these ALPs generates multiple $e^\pm$ pairs, producing positrons with energies significantly below the DM particle mass. This mechanism relaxes the constraints from Galactic diffuse $\gamma$-ray emission and expands the allowed DM mass range to above 10 MeV, a threshold necessary to avoid tension with cosmological observations from BBN and CMB. While DM annihilations directly into $e^+e^-$ pairs are insufficient to account for the observed 511 keV line and the aforementioned constraints, our model provides a natural explanation for the observed 511 keV emission without requiring fine-tuning. Importantly, from a particle-physics standpoint, the model is highly predictive: its phenomenology is effectively governed by a single parameter that simultaneously fixes the thermal relic abundance and the annihilation rate relevant for the 511 keV line. This minimality constitutes a non-trivial strength of the framework. The confrontation with data, however, unavoidably depends on astrophysical inputs, most notably the DM density profile. Within these uncertainties, we find that the model is compatible with the morphology and the bulge-to-disk emission of the observed 511 keV line for a DM distribution close to a NFW profile.

Interestingly, as detailed in Appendix~\ref{sec:IfA_App}, this DM scenario predicts a significant continuum MeV emission, which could potentially explain the observed excess in MeV \(\gamma\)-rays~\cite{Karwin_2023, Orlando_2017}. Additionally, the model is testable with the upcoming CMB-S4 experiment~\cite{CMB-S4}, offering a unique opportunity to confirm or challenge its validity.

\medskip
 \emph{Acknowledgements---} 
We would like to thank Chris Capiello and Filippo Sala for useful discussions.
P.D.L. is supported by the Juan de la Cierva JDC2022-048916-I grant, funded by MCIU/AEI/10.13039/501100011033 European Union "NextGenerationEU"/PRTR. The work of P.D.L. is also supported by the grants PID2021-125331NB-I00 and CEX2020-001007-S, both funded by MCIN/AEI/10.13039/501100011033 and by ``ERDF A way of making Europe''. P.D.L. also acknowledges the MultiDark Network, ref. RED2022-134411-T. This project used computing resources from the Swedish National Infrastructure for Computing (SNIC) under project No.2022/3-27 partially funded by the Swedish Research Council through grant no. 2018-05973. D.G.~acknowledges support from the project ``Theoretical Astroparticle Physics (TAsP)'' funded by INFN. The research conducted by M.A., A.D., G.M and P.P.~receives partial funding from the European Union–Next generation EU (through Progetti di Ricerca di Interesse Nazionale (PRIN) Grant No. 202289JEW4).


\newpage	
\widetext
\setcounter{equation}{0}
\setcounter{figure}{0}
\setcounter{table}{0}
\makeatletter
\renewcommand{\theequation}{A\arabic{equation}}
\renewcommand{\thefigure}{A\arabic{figure}}


\appendix

\renewcommand{\theequation}{\Alph{section}\arabic{equation}}
\makeatletter
\@addtoreset{equation}{section}
\makeatother

\renewcommand{\thefigure}{\Alph{section}\arabic{figure}}
\makeatletter
\@addtoreset{figure}{section}
\makeatother

\renewcommand{\thetable}{\Alph{section}\arabic{table}}
\makeatletter
\@addtoreset{table}{section}
\makeatother

\section{Dark Matter annihilation}\label{sec:ann}
\paragraph*{\bf Total cross sections:} As outlined in the main text, the model under consideration extends the Standard Model (SM) by introducing an axion-like particle (ALP) $a$ that couples to all three SM leptons, $\ell = e, \mu, \tau$, as well as to a Dirac dark matter (DM) particle $\chi$. The interaction Lagrangian for this model is provided in Eq.~\eqref{eq:lagrangian} of the main text. 

\begin{figure}[h!]
    \begin{center}
    \includegraphics[width=0.6\linewidth, trim={0 5cm 0 5cm}, clip]{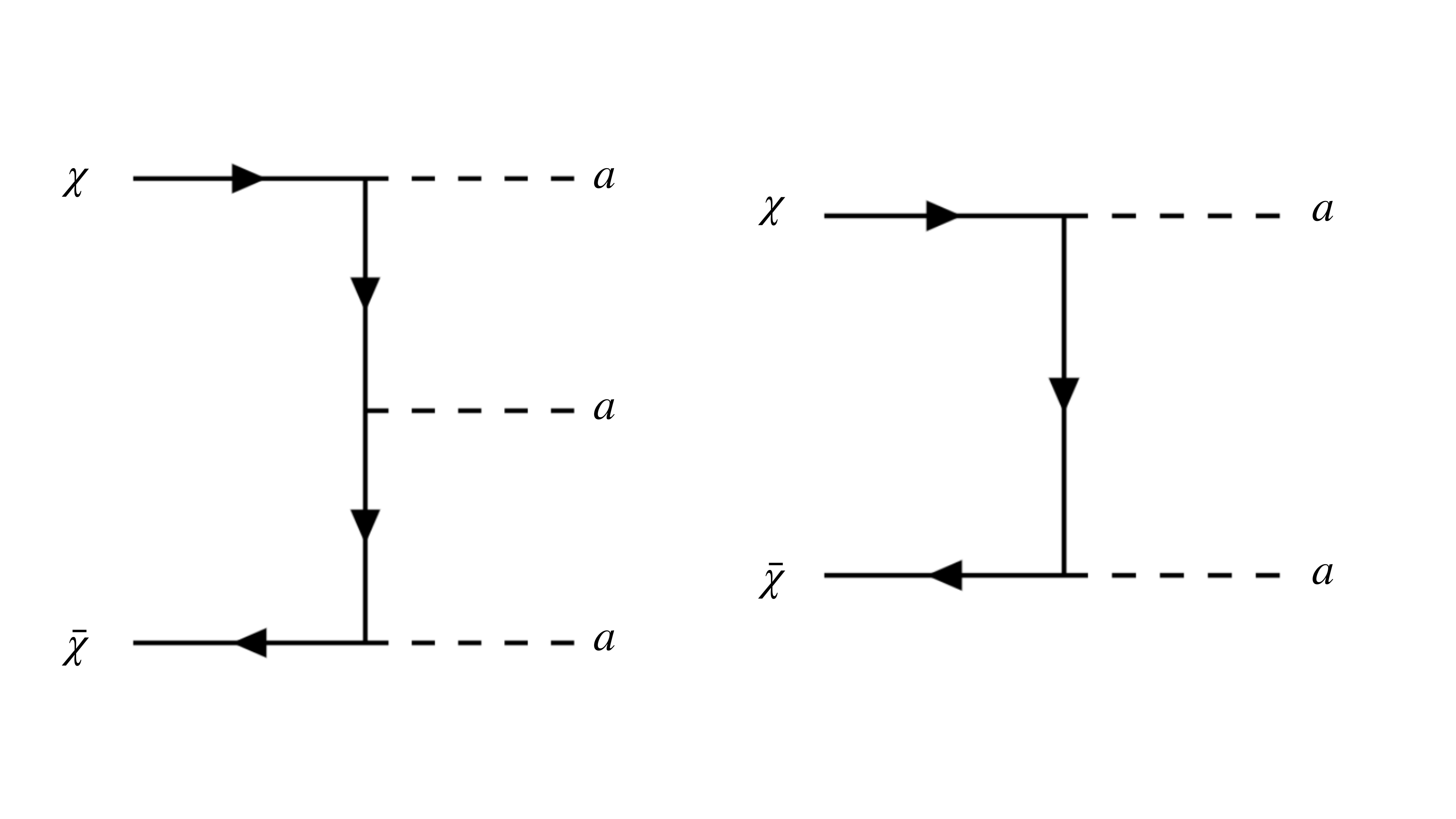}
    \end{center}
    
    \caption{Feynman diagrams for the \textit{s-}wave annihilation into three pseudoscalars (\textit{left panel}), and the \textit{p-}wave into a pair of pseudoscalars (\textit{right panel}).}
    \label{fig:feyn_diagrams}
\end{figure}

Within the mass hierarchy $2m_\chi > 3m_a > 6m_e$ and for $m_\chi < 100\,\text{MeV}$, DM particles can annihilate into ALPs, which subsequently decay exclusively into $e^\pm$ pairs at tree level. The ALP-DM coupling, $g_\chi$, fully governs the annihilation processes, while the electron coupling, $g_e$, primarily determines the ALP mean free path. Although $g_e$ is tightly constrained by multiple bounds, a viable parameter space exists where the mean free path on galactic scales is sufficiently small (less than the parsec scale), while the cross-section for direct annihilation into $e^\pm$ via off-shell ALP mediation remains negligible.
As a result annihilation proceeds through the $p$-wave $\chi\bar{\chi} \to aa$ and the $s$-wave $\chi\bar{\chi} \to aaa$ processes. The Feynman diagrams for both t-channel processes are illustrated in Fig.~\ref{fig:feyn_diagrams}.

Regarding the \textit{p-}wave process, in the non-relativistic limit, where $s \simeq m_\chi^2\left(4 + \bar{v}^2\right)$ and $\bar{v}$ is the DM relative velocity, the cross section can be expressed as~\cite{Arina:2014yna, Armando_2024}:
\begin{equation}
    \sigma v_{\chi \chi\to aa}  = \frac{g_\chi^4 \bar v^2}{24\pi} 
    \frac{m_\chi^2(m_\chi^2 - m_a^2)^2}{(2m_\chi^2 - m_a^2)^4} 
    \sqrt{1 - \frac{m_a^2}{m_\chi^2}} \ .
    \label{eq:Suppxsecaa}
\end{equation}
From~\cite{Bell_2017}, the full expression for the differential cross section of the $s$-wave process can be written as:
\begin{equation}
        \frac{d  \sigma v_{\chi \chi\to aaa}}{d x_2} = \frac{g_\chi^6}{2^5 \cdot 3! \, \pi^3}
        \int_{x_1^{\rm min}}^{x_1^{\rm max}} dx_1 \, \xi(x_1, x_2, \rho_a) \ ,
    \label{eq:Suppdiffxsecaaa}
\end{equation}
where we defined the function $\xi$ as
\begin{equation}
    \xi(x_1, x_2, \rho_a) = \frac{
        \left(1 - 4(x_1 + x_2) + 4x_1x_2 + 4(x_1^2 + x_2^2) 
        + 2\rho_a^2 - 3\rho_a^4\right)^2
    }{
        8m_\chi^2 (x_1 - \rho_a^2)^2 (x_2 - \rho_a^2)^2 (1 - x_1 - x_2 - \rho_a^2)^2
    } \ .
\end{equation}

Here, $\xi$ depends on $x_{1,2} = E_{1,2} / 2m_\chi$, where $E_{1,2}$ are the energies of two of the three pseudoscalars in the center-of-mass frame of the DM pair, and $\rho_a = m_a /(2m_\chi)$ is constrained by the kinematics to be within $\rho_a\in(m_e/m_\chi,1/3)$. Note that the integrand is symmetric under the exchange of $x_1$ and $x_2$, and the integration limits are given by:
\begin{equation}
    x_{1}^{\rm min, max} = \frac{1 + 2x_2^2 + \rho_a^2 - x_2(3 + \rho_a^2) \mp A}{2(1 - 2x_2 + \rho_a^2)} \ ,
\end{equation}
where
\begin{equation}
    A = \bigg[4x_2^4 - 4x_2(x_2^2 - \rho_a^2)(1 - \rho_a^2) \quad + x_2^2(1 - 6\rho_a^2 - 3\rho_a^4) - \rho_a^2(1 - 2\rho_a^2 - 3\rho_a^4)\bigg]^{1/2} \ .
\end{equation}

To obtain the total cross-section, Eq.~\eqref{eq:Suppdiffxsecaaa} must be integrated over the range of $x_2$ allowed by kinematics:
\begin{equation}
        \sigma v_{\chi \chi\to aaa} = \int_{\rho_a}^{(1-\rho_a)/2} \frac{d\sigma v_{\chi \chi\to aaa}}{d x_2} dx_2 =  \frac{g_\chi^6}{2^5 \cdot 3! \, \pi^3}
        \int_{\rho_a}^{(1-\rho_a)/2} dx_2 \int_{x_1^{\rm min}}^{x_1^{\rm max}} dx_1 \, \xi(x_1, x_2, \rho_a) \ .
    \label{eq:Suppxsecaaa}
\end{equation}

 Remarkably, in the massless limit $m_a\to 0$, Eq.~\eqref{eq:Suppxsecaaa} simplifies to the cross section given in~\cite{Kahlhoefer_2017} and in the main text:
\begin{equation}
     \sigma v_{\chi\chi\to aaa} \simeq \frac{(7\pi^2 - 60) g_\chi^6}{1536 \pi^3 m_\chi^2} \ .
\end{equation}

\paragraph*{\bf Positron spectrum:} As outlined in the previous section and in the main text, the ALP mean-free path is smaller than the parsec scale for DM masses in the MeV to tens of MeV range, ensuring an effectively instantaneous injection of $e^\pm$ from the perspective of the 511 keV line signal. Consequently, the differential cross section for the injected positrons can be determined by convolving Eq.~\eqref{eq:Suppdiffxsecaaa} with the $e^\pm$ energy spectrum arising from the two-body decay of each ALP. We approximate the $e^\pm$ spectrum using a Dirac delta function centered at $E_2/2$. This accurately reflects the expected injection spectrum, as final state radiation is significantly suppressed at such low $e^\pm$ energies. As a result, the differential cross section as a function of the positron energy is given by
\begin{equation}
\begin{split}
    \frac{d\sigma v_{\chi\chi\to aaa}}{dE_{e}}  &= \int d x_2 \frac{d  \sigma v_{\chi \chi\to aaa}}{d x_2} \, \delta(E_e-m_\chi x_2)
  = \left. \frac1{m_\chi} \frac{d  \sigma v_{\chi \chi\to aaa}}{d x_2} \right|_{x_2 = E_e/m_\chi}=\\
  &=\frac{1}{m_\chi}\frac{g_\chi^6}{2^5 \cdot 3! \, \pi^3}
        \int_{x_1^{\rm min}}^{x_1^{\rm max}} dx_1 \, \xi\bigg(x_1, \frac{E_e}{m_\chi}, \rho_a\bigg) \ .
\label{eq:SuppdiffxsecEe}
\end{split}
\end{equation}

The number of positrons per annihilation event and unit energy is then obtained by dividing Eq.~\eqref{eq:SuppdiffxsecEe} by the total cross section Eq.~\eqref{eq:Suppxsecaaa}:
\begin{equation}
    \frac{d{\cal N}_e}{dE_e}\equiv \frac{1}{\sigma v_{\chi\chi\to aaa}}\frac{d\sigma v_{\chi\chi\to aaa}}{dE_{e}} = \frac{1}{m_\chi}\frac{\int_{x_1^{\rm min}}^{x_1^{\rm max}} dx_1 \, \xi\bigg(x_1, \frac{E_e}{m_\chi}, \rho_a\bigg)}{\int_{\rho_a}^{(1-\rho_a)/2} dx_2 \int_{x_1^{\rm min}}^{x_1^{\rm max}} dx_1 \, \xi(x_1, x_2, \rho_a)} \ .
    \label{eq:SuppdNdE}
\end{equation}

As far as the DM spatial distribution $\rho_\chi(\vec{x})$ is concerned, we consider hereafter a generalized NFW profile. 
The $e^\pm$ injection rate per unit volume and energy resulting from the DM annihilation process can be expressed by the source term:
\begin{equation}
    Q_e(\vec{x},E_e)=\frac{\langle\sigma v\rangle_{\chi\chi \to aaa}}{4}\bigg(\frac{\rho_\chi(\vec{x})}{m_\chi}\bigg)^2\frac{d{\cal N}_e}{dE_e} \equiv  \frac{1}{4}\bigg(\frac{\rho_\chi(\vec{x})}{m_\chi}\bigg)^2 \left. \frac{1}{m_\chi} \frac{d \sigma v_{\chi \chi\to aaa}}{d x_2} \right|_{x_2 = \frac{E_e}{m_\chi}} \ ,
\label{eq:Suppsource_term}
\end{equation}
where the extra factor of \(1/2\) compared to the standard source term for self-conjugate DM reflects the Dirac nature of the DM particle. Moreover, for \(s\)-wave processes in the non-relativistic regime, the thermally averaged cross-section is simply given by \(\langle\sigma v\rangle_{\chi\chi \to aaa} = \sigma v_{\chi\chi \to aaa}\).

\section{Relic Density}\label{sec:relic density}
Following the approach outlined in~\cite{Servant_2003}, we solve the Boltzmann equation for thermal freeze-out, considering only the dominant \(p\)-wave annihilation cross section $\sigma v_{\chi\chi\to aa}$. The comoving number density is defined as \(Y = n_\chi/s\), where \(n_\chi\)  is the DM number density, and  \(s\) is the entropy density. 
As discussed in the main text, two temperatures play an important role: $T$, the temperature of the SM plasma, and $T'$, the temperature of the dark sector. We define the variables $x = m_\chi / T$ and $x' = m_\chi / T'$, with their ratio given by $x/x'=(g_\star(T)/g_\star(T_{a\rm SM}))^{1/3}$ as already noted in the main text. The entropy density and the Hubble rate take their usual forms, $s(x)=s(1)x^{-3}$ and $H(x)=H(1)x^{-2}$ respectively. As a consequence, the Boltzmann equation becomes: 
\begin{equation}\label{eq:SuppBE_ann}
		\frac{dY}{dx} = -\frac{\lambda \left<\sigma v\right>_{\chi\chi\to aa}}{x^2}(Y^2 - Y^2_{\rm eq}(x')) \ ,
\end{equation}  
where $\lambda = s(1)/H(1)$ and the thermally averaged annihilation cross section $\langle \sigma v\rangle_{\chi\chi\to aa}$ is evaluated by substituting $\bar{v}^{\,2} = 6/x'$ into Eq.~\eqref{eq:Suppxsecaa}. Here, $Y_{\rm eq}(x')$ denotes the equilibrium comoving number density of the dark sector species.  
Following the procedure of~\cite{Servant_2003}, for a non relativistic relic the solution of Eq.~\eqref{eq:SuppBE_ann} can be approximated analytically, yielding the freeze out abundance:
\begin{equation}\label{eq:SuppYinf}
	Y_{\infty} \simeq \bigg(\frac{g^\star_s(T_{a\rm SM})}{g^\star_s}\bigg)^{1/3}\sqrt{\frac{45}{\pi}} \frac{\sqrt{g_{\rho}^*}}{g^*_s} \frac{1}{m_{\chi} M_{\rm pl}} \left( \frac{x_{\rm fo}'}{\left<\sigma v\right>_{\chi\chi\to aa} (2 x_{\rm fo}')} \right) \ ,
\end{equation}  
where $g^\star_{\rho}$ and $g^\star_s$ are the SM relativistic degrees of freedom for energy density and entropy, respectively, evaluated at the dark sector temperature $T'$ and $M_{\rm pl}$ is the Planck mass. The only modification with respect to the standard freeze-out scenario is the ratio of degrees of freedom at SM–dark sector decoupling and at freeze out, as emphasized in~\cite{Dror:2023fyd}. The freeze-out point is determined solely by the dark sector dynamics, and from the condition $\Gamma(x'_{\rm fo}) \simeq H(x'_{\rm fo})$ one obtains the iterative solution:
\begin{equation}\label{eq:Suppzfo}
	x_{\rm fo}' = \log{\left[\zeta(\zeta + 2) \sqrt{\frac{45}{8}} \frac{g_{\rm DM}}{2\pi^3} \frac{m_{\chi} M_{\rm pl} \left<\sigma v\right>_{\chi\chi\to aa} (x_{\rm fo}')}{\sqrt{g^*_s x_{\rm fo}'}}\right]} \ ,
\end{equation}  
with the choice \(\zeta = 1/2\) providing a good match to the full numerical solution and \(g_{\rm DM} = 4\) for a Dirac fermion.
Here we consider the scenario in which $T_{a\rm SM} \gg v_{\rm EW}$, where $v_{\rm EW}$ is the electroweak vacuum expectation value. Since we focus on dark matter masses of order $\mathcal{O}(10~\mathrm{MeV})$, we take $g_s^\star = g_\rho^\star \simeq 10.75$, which fixes the ratio $\left( g_s^\star(T_{a\rm SM})/g_s^\star(x_{\rm fo}) \right)^{1/3} \simeq 2.15 .$

\smallskip
The contribution of \(\chi\) to the Universe's energy density is expressed as \(\Omega = \rho_\chi / \rho_{\rm c}\), where \(\rho_{\rm c}\) is the critical density corresponding to a flat universe. The DM energy density, in terms of the freeze-out abundance, is given by \(\rho_\chi = m_\chi s_0 Y_\infty\), with \(s_0\) representing the present-day entropy density. By requiring \(\Omega h^2 \simeq 0.12\)~\cite{Planck2018}, the coupling coefficient \(g_\chi\) can be determined as a function of the DM mass \(m_\chi\). We approximately obtain $g_\chi \simeq 2.5 \times 10^{-3} \left( {m_\chi}/{\text{MeV}} \right)^{1/2}$ with only a very small dependence on $\rho_a$ within the parameter space explored in this study.

\medskip
Using these results, the total \(s\)-wave cross-section is computed. Table~\ref{tab:xsectab} presents the results for different values of \(\rho_a\) and \(m_\chi\) expressed in $\si{cm^3\,s^{-1}}$ units. The white cells correspond to kinematically forbidden regions.
 \begin{table}[t!]
    \centering
    \renewcommand{\arraystretch}{1.5}
    \setlength{\tabcolsep}{10pt}
    \definecolor{pastelblue}{RGB}{173, 216, 230}
    \definecolor{customgreen}{RGB}{204, 255, 204}

    \caption{
        Cross-section values $\langle \sigma v \rangle_{\chi\chi\to aaa}$ 
        as a function of $\rho_a$ and $m_\chi$, expressed in $\mathrm{cm^3\,s^{-1}}$.
    }
    \label{tab:xsectab}

    \begin{tabular}{|c||c|c|c|c|c|}
    \hline
     & \cellcolor{pastelblue}\textbf{5 MeV}
     & \cellcolor{pastelblue}\textbf{7 MeV}
     & \cellcolor{pastelblue}\textbf{10 MeV}
     & \cellcolor{pastelblue}\textbf{15 MeV}
     & \cellcolor{pastelblue}\textbf{20 MeV} \\
    \hline\hline

    \cellcolor{customgreen}$\rho_a = 1/4$ 
        & $1.48\times10^{-31}$ 
        & $1.82\times10^{-31}$
        & $2.34\times10^{-31}$
        & $3.07\times10^{-31}$
        & $3.74\times10^{-31}$ \\
    \hline

    \cellcolor{customgreen}$\rho_a = 1/8$ 
        & $4.30\times10^{-31}$ 
        & $5.27\times10^{-31}$
        & $6.78\times10^{-31}$
        & $8.88\times10^{-31}$
        & $1.09\times10^{-30}$ \\
    \hline

    \cellcolor{customgreen}$\rho_a = 1/10$ 
        & -- 
        & $5.90\times10^{-31}$
        & $7.59\times10^{-31}$
        & $9.95\times10^{-31}$
        & $1.22\times10^{-30}$ \\
    \hline

    \cellcolor{customgreen}$\rho_a = 1/12$ 
        & -- 
        & $6.30\times10^{-31}$
        & $8.00\times10^{-31}$
        & $1.06\times10^{-30}$
        & $1.30\times10^{-30}$ \\
    \hline

    \cellcolor{customgreen}$\rho_a = 1/16$ 
        & -- 
        & -- 
        & $8.62\times10^{-31}$
        & $1.15\times10^{-30}$
        & $1.40\times10^{-30}$ \\
    \hline

    \end{tabular}
\end{table}

\section{In-flight annihilation emission}
\label{sec:IfA_App}

We evaluate the in-flight positron annihilation (IfA) signals using the approach outlined in Ref.~\cite{Beacom_2006}, generalized to account for the fact that the \(e^+\) injected from DM annihilations are not monochromatic but follow the spectrum described by Eq.~\eqref{eq:SuppdNdE}. This prescription is employed as it enables a model-independent estimation of the IfA emission, independent of propagation modeling, in a specific region of the Galaxy. The estimation is normalized to the 511 keV intensity emission observed in that region. With this normalization, the photon flux from IfA is given by
\begin{equation}
    \begin{split}
        & \frac{d\phi^\gamma_\textrm{IfA}}{d\Omega \,dE_{\gamma}} = \frac{d\phi^\gamma_{511 \, \rm keV}}{d\Omega} \frac{n_{\rm H}}{P(1-\frac{3}{4}f)} \int^{E_\textrm{max}}_{E_{\gamma}} dE'\,\frac{1}{\mathcal{N}_{e}} \frac{d\mathcal{N}_{e}}{dE'} \int^{E'} _{m_e}P_{E'\rightarrow E}\, \frac{d\sigma}{dE_{\gamma}} \frac{dE}{|dE/dx|} \ ,
    \end{split}
    \label{eq:SuppSuppIA}
\end{equation}
which also depends on the electron-positron annihilation cross sections~\cite{Dirac} (assuming a warm ionized gas) and the energy loss rate. Specifically, the second integral over \(E'\) accounts for the energy distribution of positrons injected by DM, as given in Eq.~\eqref{eq:SuppdNdE}, where \(E_\textrm{max} = \left(1-\rho_a\right)/2 \, m_\chi \) and \(E_\textrm{min}\) represents the minimum positron energy required to produce \(\gamma\)-rays with energy \(E_\gamma\). The term \((1/\mathcal{N}_e) \, (d\mathcal{N}_e/dE')\)  represents the fraction of positrons at energy \(E'\) contributing to the 511 keV emission. The other terms describe the energy loss for a positron with energy \(E'\) before annihilation, the number density of electrons \(n_{\rm H}\) including both bound electrons within hydrogen atoms and free electrons (which determine the number of targets for positron scattering), and the probability \(P_{E'\to E}\) (from Eq.~(4) of Ref.~\cite{Beacom_2006}), which gives the likelihood per unit energy for a positron with initial energy \(E'\) to produce a \(\gamma\)-ray in flight before reaching energy \(E\). Additionally, \(P = P_{E'\to m_e}\) represents the probability for a positron with initial energy \(E'\) to produce a photon before thermalizing, and \(f = 0.967 \pm 0.022\)~\cite{Jean_2005} is the fraction of positrons annihilating through positronium states.

\begin{figure*}[t!]
    \centering
\includegraphics[width=0.49\linewidth]{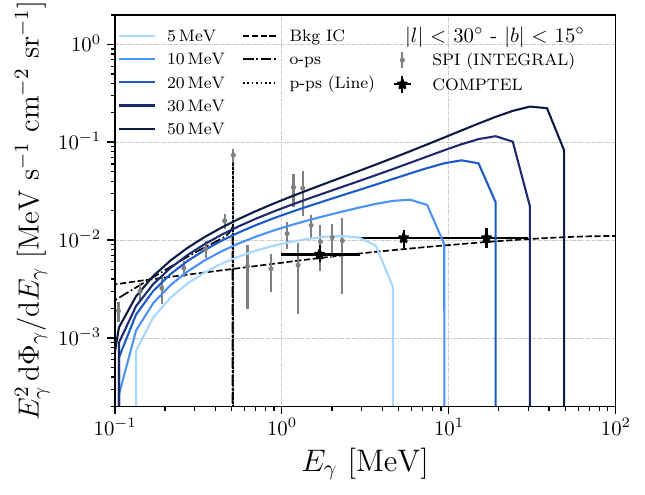}
\includegraphics[width=0.49\linewidth]{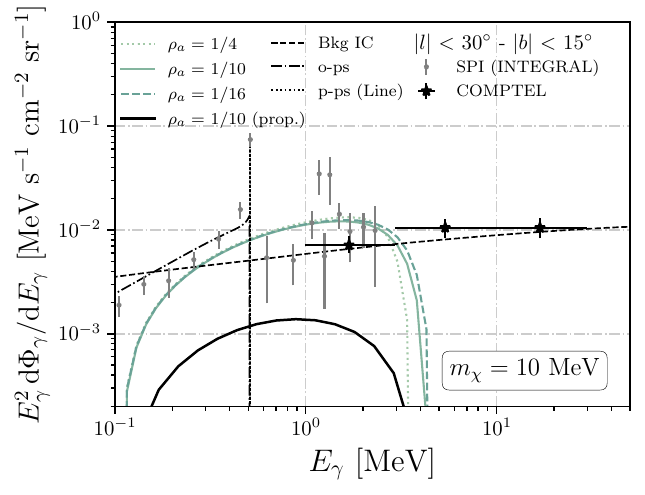}
\caption{
\textbf{Left panel:} Predicted IfA signals from DM annihilating in the conventional scenario (i.e. injecting monoenergetic positrons), for different DM masses. \textbf{Right panel:} Expected in-flight annihilation signal for a $10$~MeV DM particle for different values of $\rho_a$. The IfA emission is computed following the general and model-independent prescription of Ref.~\cite{Beacom_2006}, which directly relates the signal to the injected positron spectrum and provides a conservative estimate. This choice contrasts with the approach of Ref.~\cite{DelaTorreLuque:2024wfz}, where the IfA emission is evaluated from the steady-state positron distribution after propagation effects such as energy losses and reacceleration. For reference, we also show the IfA signal obtained by including full positron propagation for the benchmark case $\rho_a = 1/10$ and contraction index $\gamma = 1.3$, using the same transport setup adopted for the $511$~keV line, explicitly illustrating that our DM model always leads to a lower IfA emission.}
    \label{fig:IfA_comparisons}
\end{figure*}

\medskip
In the region of interest of SPI and COMPTEL data (\textit{i.e.} latitude $|b|<15^{\circ}$ and longitude $|l|<30^{\circ}$), the line peak flux is $d\phi^\gamma_{511\, \rm keV}/d\Omega = 0.07$~MeV cm$^{-2}$ s$^{-1}$ sr$^{-1}$, as can be seen in Fig.~\ref{fig:IfA_comparisons}. In the left panel of this figure, we present the IfA continuum signals from the annihilation of conventional DM (direct annihilation into an $e^\pm$ pair) for different DM mass values.  In the right panel, we show the IfA emission predicted by our model, fixing the DM mass  at $m_\chi = 10$ MeV and varying $\rho_a$. Here, it is evident that IfA emissions are not significantly affected by variations in $\rho_a$. This behavior is also observed for other choices of DM mass. In both panels, the intensities of the 511~keV line and the associated ortho-positronium continuum emissions are shown as dotted and dot-dashed black lines, respectively, while the signal from the IC background is represented by a dashed black line.

\medskip
The comparison between the left panel of Fig.~\ref{fig:IfA_comparisons} and Fig.~\ref{fig:IfA_MassScan} in the main text highlights the improvement of our model over the conventional DM scenario, which already forbids a DM mass of 10~MeV. Conversely, the IfA emission predicted by our model is compatible with COMPTEL and SPI data.

We also remark that models of CR electrons tuned on Voyager-1 data, fail to reproduce the $\gamma-$ray emission in the inner Galaxy at MeV energies. Remarkably, the leptons produced within our scenario could help to explain the apparent excess~\cite{Karwin_2023, Orlando_2017}. However, this anomaly can be avoided considering models~\cite{delaTorreLuque:2022vhm, DelaTorreLuque:2023zyd}, that are only constrained by AMS-02 data and Fermi-LAT local emissivity measurements.

\section{Line morphology: Numerical Setup and Parameter Variations}
\label{sec:Line_App}

In order to obtain the line signal, we need to calculate numerically the distribution of the positrons, originating from DM annihilation, after propagation in the interstellar medium. 
To this aim, we use the {\tt DRAGON} code \cite{Evoli:2016xgn}, a numerical package designed to compute all the physical processes associated to the transport of charged particles in the Galaxy including diffusion, stochastic reacceleration, advection, spallation, and energy losses. 

Regarding the latter, which is the most relevant for MeV electrons/positrons, {\tt DRAGON} can realistically take into account inverse Compton scattering, synchrotron emission, bremsstrahlung, ionization and Coulomb scattering, embedding up-to-date models for the distribution of the astrophysical targets of interest for each process. 
We remark that, in the energy domain we are focusing on, the timescales associated to Coulomb and ionization losses are by far the dominant ones (see for instance the discussion in \cite{Bartels:2017dpb} for more details): hence, the large uncertainties due to the details of diffusive transport have little impact on our final result.

Once all the relevant processes are correctly initialized, the injection of charged leptons associated to our DM model is implemented as in Eq. \ref{eq:Suppsource_term}.
The charged particles are then propagated in a two-dimensional cylindrically symmetric spatial grid with spacing $\simeq 100$ pc, which is appropriate to resolve the spatial variations of the energy loss term. The kinetic energy range under consideration is [$100$ eV - $100$ MeV]. The convergence is reached adopting a variable timestep with a maximum value of $\delta t = 64$ Myr and a minimum value of $\delta t = 100$ y, which is appropriate to cover the different timescale involved in the problem.
We adopt the propagation parameters discussed in Ref.~\cite{DelaTorreLuque:2023olp}, derived from state-of-the-art analyses of secondary cosmic rays in the Galaxy. 
The runs are available at \url{https://doi.org/10.5281/zenodo.10076728}.

With the propagated flux of charged leptons at hand, the 511~keV line flux integrated along the line of sight (where the variable \(s\) runs) is expressed as follows (for more details, see Ref.~\cite{DelaTorreLuque:2024wfz}):
\begin{equation}
    \frac{d\phi^\gamma_{511\, \rm keV}}{d\Omega}=6k_{\rm ps} \int_{\rm l.o.s.}  \hspace{-.1cm} ds \, \left( \frac{d\phi^{e^+}}{d\Omega}\right)_{\rm propagated} \hspace{-.6cm} n_e \cdot \sigma_\textrm{ann}(E_\textrm{th})\ ,
    \label{eq:Supp511line}
\end{equation}
where: $k_{\rm ps}=1/4$ is the fraction of positronium decays contributing to the $511$~keV line signal, 
 the factor $6$ accounts for the emission of two $511$~keV photons per positron annihilation times three positrons injected per DM annihilation in the process $\chi\chi\to aaa\to 3(e^+e^-)$, $n_e$  is the free electron density, and $\sigma_\textrm{ann}$ the cross section of charge exchange with hydrogen~\cite{Pos_transport} at the energy of the thermalized positrons $E_\textrm{th}$. We consider the thermal energy  of a warm medium (with $T=8000$~K), as suggested by Ref.~\cite{Knoedlseder_2005}. See Ref.~\cite{DelaTorreLuque:2024wfz} for more details.

\begin{figure*}[t!]
    \centering
    \includegraphics[width=0.48\linewidth]{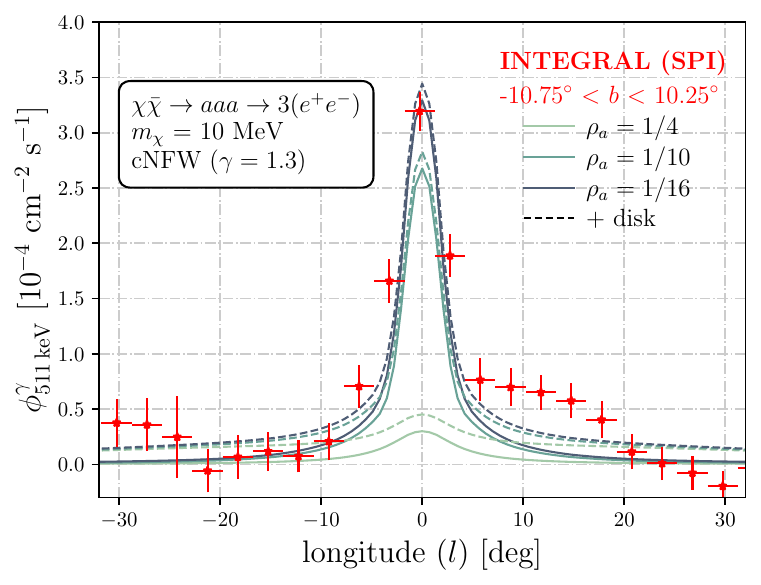} \,
        \includegraphics[width=0.48\linewidth]{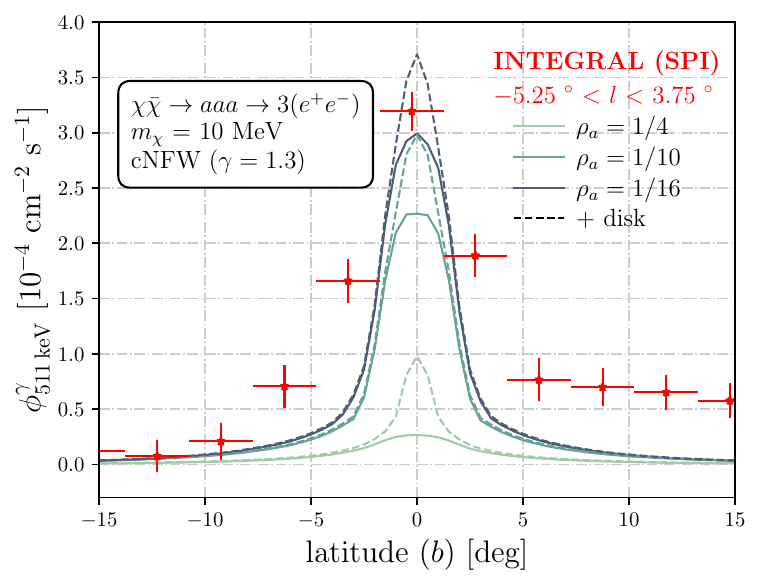}
        
    \includegraphics[width=0.48\linewidth]{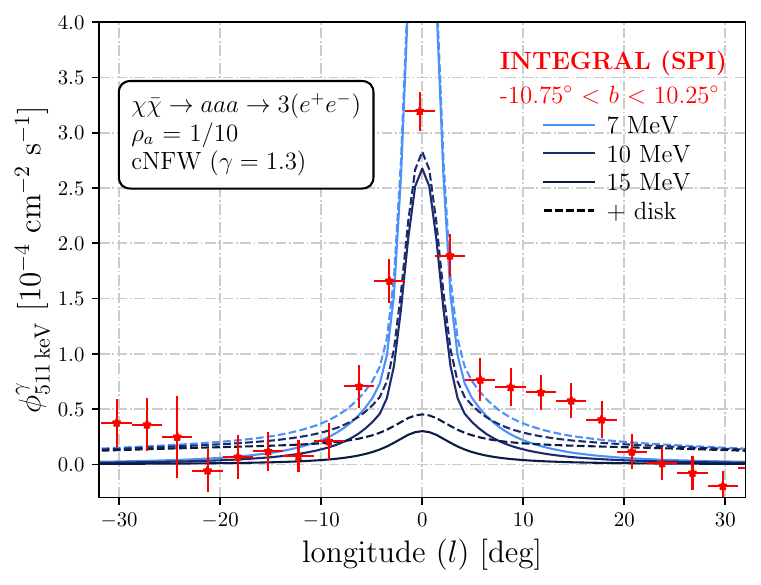} \,
    \includegraphics[width=0.48\linewidth]{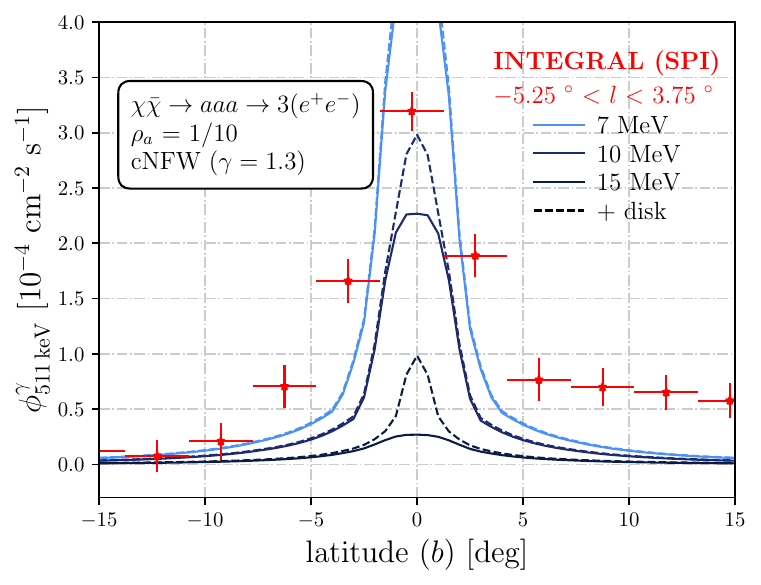}
    \caption{\textbf{Top row:} Longitude profile (\textit{left panel}) and latitude (\textit{right panel}) for different values of $\rho_a$ and a DM mass of 10 MeV, adopting an cNFW ($\gamma = 1.3$) DM distribution similar to the one reproducing the Galactic Center Excess. \textbf{Bottom row} Longitude (\textit{left panel}) and latitude (\textit{right panel}) profiles for different DM masses, at $\rho_a = 1/10$ and adopting an cNFW ($\gamma = 1.3$) DM distribution.
    }
    \label{fig:App_511keV_rhos}
\end{figure*}

\medskip
Let us now present how the results obtained within this setup vary with respect to the choice of the most relevant parameters. In the top row of Fig.~\ref{fig:App_511keV_rhos} we show the expected longitude (left panel) and latitude (right panel) profiles for different values of $\rho_a$ (keeping the DM mass fixed at 10 MeV), for a contracted NFW (c-NFW) with contraction index $\gamma = 1.3$. 
The differences in normalization arise from the direct dependence of the annihilation rate on \(\rho_a\) and the kinematics of the positrons injected (that is why varying \(\rho_a\) slightly influences the morphology of the expected signal). One can see that there are different combinations of \(\rho_a\) and the contraction index that would lead to a remarkable reproduction of the excess' morphology.  In the bottom row of the figure, we compare the longitude (left panel) and latitude (right panel) profiles predicted for different DM masses at \(\rho_a = 1/10\), assuming a contracted NFW (c-NFW) profile with a contraction index \(\gamma = 1.3\). For various parameter choices, the SPI profiles are well-matched, demonstrating the capability of the proposed model to potentially address a longstanding astrophysical puzzle, while reducing the continuum signal associated.

\bibliography{references.bib}

\end{document}